\documentclass [prd,reprint,eqsecnum,nofootinbib]{revtex4-1}
\usepackage{amsfonts,amsmath,hyperref,url,color}

\def\cD{{\cal D}}

\def\cL{{\cal L}}

\begin{document}

\title{%
Gravity as a constrained BF theory:\\ Noether charges and Immirzi
parameter}
\author{R. Durka}
\email{rdurka@ift.uni.wroc.pl}\affiliation{Institute for Theoretical Physics,
University of Wroc\l{}aw, Pl.\ Maxa Borna 9, Pl--50-204 Wroc\l{}aw, Poland}
\author{J. Kowalski-Glikman}
\email{jkowalskiglikman@ift.uni.wroc.pl}\affiliation{Institute for Theoretical Physics,
University of Wroc\l{}aw, Pl.\ Maxa Borna 9, Pl--50-204 Wroc\l{}aw, Poland}
\date{\today}
\small
\begin{abstract}\noindent
We derive and analyze  Noether charges associated with the
diffeomorphism invariance for the constrained $SO(2,3)$ BF theory.
This result generalizes the Wald approach to the case of the first
order gravity with a negative cosmological constant, the Holst
modification and topological terms (Nieh-Yan, Euler, and
Pontryagin). We show that differentiability of the action is
automatically implemented by the the structure of the constrained BF
model. Finally, we calculate the AdS--Schwarzschild black hole
entropy from the Noether charge and we find that it
does not depend on the Immirzi parameter.
\end{abstract}
\maketitle
\section{Introduction}

The Wilsonian perspective is a powerful guiding principle in
constructing theories with the given field content and symmetries.
It tells that one should include in the action all terms that can be
constructed from the fields and are compatible with the symmetries
of the theory. In the context of first order gravity we have to do
with two fields, tetrad $e^a$ and connection $\omega^{ab}$, and two symmetries, local Lorentz invariance and spacetime diffeomorphisms. If we implement
the diffeomorphism invariance, assuming that the action of gravity
is written as a four form polynomial constructed from the tetrad and the connection, the
list of possible terms turns out to be rather short and includes
\begin{itemize}
\item Palatini Lagrangian
\begin{equation}\label{1}
  \mathcal{L}_P= R^{ab}\wedge e^{c}\wedge e^{d}\,\epsilon_{abcd}\, ,
\end{equation}
\item Cosmological term
\begin{equation}\label{2}
    \mathcal{L}_{C}=e^{a}\wedge e^{b}\wedge e^{c}\wedge e^{d}\, \epsilon_{abcd}\, ,
\end{equation}
\item Holst term \cite{Holst:1995pc}
\begin{equation}\label{3}
 H_4 =  R^{ab}\wedge e_a \wedge e_b\, ,
\end{equation}
\item Pontryagin, Euler and Nieh-Yan topological terms
\begin{eqnarray}
P_4 &=& R^{ab}\wedge R_{ab}\, ,\nonumber\\
\label{4}E_4 &=& R^{ab}\wedge R^{cd} \,\epsilon_{abcd}\, ,\\
NY_4&=& T^a \wedge T_a - R^{ab}\wedge e_a \wedge e_b\, ,\nonumber
\end{eqnarray}
\end{itemize}
where $R^{ab}$ is the curvature of $\omega^{ab}$ and $T^a$ is torsion.

Each of these terms comes with its own coupling constant. One could
ask if there is an additional principle that could be used to reduce
the number of independent parameters of the theory. As it turns out,
this can be achieved in the framework of the  formulation of gravity
as a constrained BF theory.

This approach has its roots in MacDowell-Mansouri
\cite{MacDowell:1977jt}, \cite{Stelle:1979aj} and Plebanski
\cite{Plebanski:1977zz,Capovilla:1991qb,Capovilla:1991kx} theories
 and
was developed in the series of papers \cite{Starodubtsev:2003xq,Smolin:2003qu,Freidel:2005ak,Wise:2006sm}. In
this formulation we have the anti-de Sitter
algebra $so(2,3)$-valued\footnote{The de Sitter case $so(1,4)$ can
be constructed analogously. Here we use the anti-de Sitter algebra
because it leads to the asymptotically anti-de Sitter spacetimes.}
connection  $A^{IJ}$, with $I,J=0, \ldots, 4$, which can be decomposed
into Lorentz connection  $\omega^{ab}$ and the tetrad (soldering)
one-form $e^a$ ($a,b = 0, \dots,3$) as follows
\begin{equation}\label{5}
    A^{ab} = \omega^{ab}\,,\qquad  A^{a4}=\frac{1}{\ell}\, e^a\,.
\end{equation}
Here $\ell$ is a length scale necessary for dimensional reason since
the tetrad is dimensionless. As we will see this scale is naturally
associated with the cosmological constant. The components of the
curvature of connection $A^{IJ}$ are related to the curvature of
Lorentz connection $\omega$
\begin{equation}\label{6}
F^{ab}(A)= R^{ab}(\omega)+ \frac1{\ell^2}\, e^a\wedge e^b
\end{equation}
and the torsion
\begin{equation}\label{7}
    F^{a4} =\frac1{\ell}\left(d e^{a}  +
\omega^a{}_b\wedge e^b \right)=
 \frac1{\ell}\, T^a\, .
\end{equation}
With the help of the second ingredient, the $so(2,3)$ Lie algebra
valued two-from field $B^{IJ}$ one can write down the action of the
theory as follows
\begin{equation}\label{8}
 16\pi \, S(A,B)= \int F^{IJ}\wedge B_{IJ} -
 \frac{\beta}{2} B^{IJ}\wedge B_{IJ} - \frac{\alpha}{4}\epsilon^{IJKL4} B_{IJ}\wedge
 B_{KL}
\end{equation}
After solving $B$ field equations we find
\begin{equation}\label{9}
  B^{a4} = \frac1\beta\, F^{a4},
  \quad B^{ab}
  =\frac{1}{2(\alpha^2+\beta^2)}( \beta \delta^{ab}_{cd}-\alpha \epsilon^{ab}{}_{cd})\, F^{cd}\, .
\end{equation}
Before substituting this result back to the action (\ref{8}) let us
provide the expressions for dimensionless coupling constants
$\alpha$ and $\beta$ and the scale $\ell$ in terms of the physical
coupling constants, Newton's constant $G$, a negative cosmological
constant $\Lambda$, and the Immirzi parameter $\gamma$
\cite{Immirzi:1996di}
\begin{equation}\label{10}
    \alpha =
\frac{G\Lambda}{3}\frac{1}{(1+\gamma^2)}, \quad \beta =
\frac{G\Lambda}{3}\frac{\gamma}{(1+\gamma^2)} , \quad
\gamma=\frac{\beta}{\alpha}\, , \quad \Lambda = -\frac{3}{\ell^2}\,
.
\end{equation}
Substituting (\ref{9}) and (\ref{10}) to the action (\ref{8}) gives
\begin{eqnarray}
 32\pi &G&\, S=\int\left(
 R^{ab}\wedge e^{c}\wedge e^{d}
 +\frac{1}{2\ell^2}  e^{a}\wedge e^{b}\wedge e^{c}\wedge
e^{d}\right) \epsilon_{abcd}\nonumber\\
& &\qquad+\frac{2}{\gamma}\int R^{ab}\wedge e_{a}\wedge e_{b}
\label{11}\\
&&\qquad+\frac{\ell^2}{2}\,
E_4-\ell^2\gamma\, P_4+\frac{2(\gamma^2+1)}{\gamma}\, NY_4\, .\nonumber
\end{eqnarray}

The first line in (\ref{11}) is the standard first order form of the
general relativity action with the cosmological constant. The third
line contains the combination of topological invariants and
therefore can be written as a total derivative (see
\cite{Durka:2010zx}). The middle term is called the Holst term
\cite{Holst:1995pc}. Although it is not a total derivative, by the virtue of the Bianchi identity, it does not influence equations of motion when
torsion vanishes.

The first order action above can be also written down in a compact
form
\begin{equation}\label{our}
 S(\omega,e)=\frac{1}{16\pi} \int_{M}\left( \frac{1}{4}M^{abcd} F_{ab}\wedge F_{cd}-\frac{1}{\beta\ell^2} \,T^a \wedge T_a\right)\,
\end{equation}
with
\begin{equation}\label{M}
    M^{ab}{}_{cd}=\frac{\alpha}{(\alpha^2+\beta^2)}( \gamma\, \delta^{ab}_{cd}
    -\epsilon^{ab}_{\;\;cd}) \equiv -\frac{\ell^2}{G}( \gamma\, \delta^{ab}_{cd}-\epsilon^{ab}_{\;\;cd})\,.
\end{equation}
The field equations following from (\ref{our}) read
\begin{eqnarray}
K^{abcd}\, F_{ab}\wedge e_c&=&0 \label{12}\\
\frac{1}{\ell^2}\cD^\omega \left(K^{abcd}(e_a \wedge
e_b)\right)&=&0\label{13}
\end{eqnarray}
where the operator $K$ has the form
\begin{equation}\label{14}
K^{ab}{}_{cd}\equiv -\frac{\ell^2}{G}\left(\frac{1}{\gamma}\,
\delta^{ab}{}_{cd}+\epsilon^{ab}{}_{cd}\right)
\end{equation}
and we have introduced AdS curvature $F$
\begin{equation}\label{15}
    F^{ab}=(R^{ab}+\frac{1}{\ell^2}e^a\wedge e^b)\,.
\end{equation}

Later we will make use of the fact that this curvature vanishes for anti-de
Sitter spacetime.

It follows from equation (\ref{13}) that
torsion $T^a=\cD^\omega e^a$ vanishes (one has to assume that
$\gamma^2 \neq -1$) and thus the field equations (\ref{14}) are
Einstein equations with a negative cosmological constant in the first
order form.\newline

\indent The Immirzi parameter, being the coupling constant associated with
the Holst term, is a mysterious beast. It was first introduced by
Barbero \cite{Barbero:1994ap} in the context of Ashtekar variables,
parametrizing a family of canonical transformations on the gravity
phase space and inequivalent quantizations. It was soon realized
that $\gamma$ is explicitly present in the Loop Quantum Gravity
formula for area spectrum \cite{Immirzi:1996di}, \cite{Rovelli:1994ge}. As a consequence, Immirzi parameter is also
present in the formula for black hole entropy calculated by counting LQG
 microstates of isolated horizon \cite{Ashtekar:2000eq,Domagala:2004jt,Meissner:2004ju,Engle:2009vc,Engle:2010kt};
 for the recent discussion of the results  see \cite{Agullo:2010zz}. On the other
 hand, as it was said above, the inclusion of the Holst term does not lead to any modifications of classical field equations of gravity and therefore is seemingly completely irrelevant classically.

However, it is well known that black hole entropy can be computed
\cite{Wald:1993nt},  \cite{Iyer:1995kg} in a class of diffeomorphism
invariant theories as a Noether charge associated with a timelike
Killing vector with a vanishing norm at the horizon.

A natural question arises: if we calculate the black hole entropy
following  the Wald and Iyer recipe in the theory of gravity with
Holst term, will we reproduce the Loop Quantum Gravity result? This
is the main problem we would like to address in this paper.

 The BF formulation of gravity is a very convenient starting point in this  context. First, it naturally leads to the emergence of the Holst term. Second,
  the analysis of the boundary terms is particularly simple in this case.
 As we will see below, in this formulation the problem of notorious
 counterterms, that usually have to be added to the action in order
 to make it differentiable and finite, is automatically taken care
 of. Last, but not least, the calculation of Noether charges in this
 formulation is much simpler than in the case of the standard first order gravity.

 The plan of this paper is as follows. In the next section we will
 show that in a black hole, asymptotically anti-de Sitter spacetime,
 the action (\ref{8}) is differentiable. This remarkable fact can be
 understood in the complementary first order gravity formulation as
 being due to the presence of the topological invariants with the right coefficients.
 In Sec.\ III, returning to the constrained BF theory, we will construct the Noether
 charges following the construction of Wald and Iyer \cite{Wald:1993nt}, \cite{Iyer:1995kg}.
 Next, in Sec.\ IV we make use of these
 expressions to calculate entropy of Schwarzschild--AdS black hole.
 The final section will be devoted to discussion and conclusions.

\section{Boundaries and differentiability}

When spacetime has boundaries we must make sure that the action (\ref{8}) is
differentiable and the variational principle is well
defined\footnote{Usually one also assumes that the action should be
finite for physically reasonable asymptotic conditions for the
fields at infinity, so as to make the path integral meaningful. We
will not investigate this issue in details here.}. The
differentiability of the action means essentially that the values of
the fields and the form of variations are chosen in such a way that
the boundary contribution to the variation of the action vanishes.
Investigating this we will see how powerful is the BF formulation
outlined in the previous section. In what follows we will restrict
ourselves to the black hole spacetimes with the anti-de Sitter
asymptotic; therefore we will have to do with a manifold with the
boundary at infinity, where the gravitational field satisfies
$F^{ab} =0$ (cf.\ \ref{15}), and the inner black hole boundary,
where we assume that the variation of connection vanishes
$\delta\omega^{ab}=0$. This latter condition is imposed because
fixing connection at the horizon means fixing the black hole
temperature, and therefore this boundary condition is essentially
equivalent to imposing the zeroth law of black hole mechanics.

Consider the variation of the action  (\ref{8}) keeping only the
terms that contribute to the boundary integral
$$
16\pi \,\delta S(A,B)= \int_M \delta F^{IJ}\wedge B_{IJ} + \ldots =
\int_M d \delta A^{IJ}\wedge B_{IJ} + \ldots
$$
\begin{equation}\label{16}
=    \int_{\partial M}\delta A^{IJ}\wedge B_{IJ}+\mbox{ bulk
    terms,}
\end{equation}
with $\partial M=(\mathbb{R}\times \partial\Sigma_{\infty}) \cup (\mathbb{R}\times \partial\Sigma_{H})$.

There are two contributions to the integral at infinity,
proportional to $B_{a4}$ and $B_{ab}$. The first vanishes because
$B_{a4}$ is proportional to torsion which vanishes by the field
equations, and the second is zero because $B_{ab} \sim F_{ab}$ which
vanishes by the virtue of asymptotic condition.

Similarly, at the black hole horizon the term $\delta A^{a4}\wedge
B_{a4}$ is proportional to torsion and therefore zero, while the
term $\delta A^{ab}\wedge B_{ab}$ vanishes because we choose the
boundary condition $\delta\omega^{ab}=0$ there, as discussed above.
Therefore, remarkably, we find that the BF action is differentiable
without any need of adding counterterms.

To understand how this result comes about let us notice that the
action (\ref{8}) written in the components  has the form\footnote{The prefactor $\gamma$ results from combining the original Holst term with torsionless part of Nieh-Yan term.}
$$
\left[(\mbox{Palatini}+\Lambda)+\ell^2 \mbox{Euler}\right]
- \gamma \left[\mbox{Holst}+\ell^2 \mbox{Pontryagin}\right]
$$
It can be checked that these are exactly the combinations needed to
cancel out the boundary terms at infinity resulting from varying the
Palatini and Holst actions. To see this consider the first
combination above. Take an arbitrary variation of the Palatini
action, to wit
\begin{equation}
\delta (\mbox{Palatini}+\Lambda)=\int_{M}(f.e.)_{a}\delta
e^a+(f.e.)_{ab}\,\delta \omega ^{ab}+\int_{\partial M}\Theta ,
\end{equation}
where $(f.e.)$ denote field (Einstein and torsion) equations, while
\begin{equation}\label{theta}
\Theta =\frac{1}{32\pi G}\epsilon_{abcd}\,\delta \omega ^{ab}\wedge
e^{c}\wedge e^{d}\,.
\end{equation}
Let us now turn to the Euler term. As it is well known
\begin{equation}
    E_4 = 32\pi\, \chi(M) + 2\int_{\partial M} \widetilde{CS}_3\, ,
\end{equation}
where $\chi(M)$ is the Euler characteristics of the manifold $M$ and
$\widetilde{CS}_3$ is the Chern--Simons three-form for the Lorentz gauge
algebra. The Euler characteristics is a fixed number and its
variation vanishes; the variation of  Chern--Simons form is
\begin{equation}\label{CS}
\delta\, \widetilde{CS}_3=\epsilon_{abcd}\,\delta \omega ^{ab}\wedge R^{cd}\,.
\end{equation}
It can be now checked directly that the terms (\ref{theta}) and
(\ref{CS}) are being combined to give $\delta\omega ^{ab}\wedge
F^{cd}\, \epsilon_{abcd}$, which is zero by the virtue of the
asymptotic condition at infinity, and by boundary condition at the
horizon. The Holst term and the Pontryagin counterterm can be
analyzed similarly.

\section{Noether charges and entropy}

Now knowing that  the action (\ref{8}) is differentiable we can turn
to the discussion of the Noether charges associated with its
symmetries. In our derivation below we will follow the procedure
proposed in the papers \cite{Wald:1993nt} and \cite{Iyer:1995kg}.
Let us start with an arbitrary variation of the action (\ref{8})
$$  \begin{aligned}
\displaystyle 16\pi\,\delta S= \int \Big(&\delta B^{IJ}
\wedge (F_{IJ}-\beta B_{IJ}-\frac{\alpha}{2}B^{KL} \,\epsilon_{IJKL4})+\\
&+\delta A_{IJ} \wedge (\cD^A B^{IJ} ) +d( B^{IJ} \wedge \delta
A_{IJ})\Big)\,.
 \end{aligned}$$
The expressions proportional to the variations of $B^{IJ}$ and
$A^{IJ}$ in the bulk are field equations, while the last term is the total
derivative of the 3-form symplectic potential:
\begin{equation}\label{17}
   \Theta=  B^{IJ} \wedge \delta A_{IJ}\,.
\end{equation}
For an arbitrary diffeomorphism generated by a smooth vector field
$\xi^\mu$, one can derive the conserved Noether current 3-form $J$
given by
\begin{equation}\label{18}
    J[\xi] = \Theta[\phi, L_\xi \phi] - I_\xi  \cL,
    \qquad  J[\xi] =B^{IJ} \wedge L_\xi A_{IJ}-I_\xi  \cL
\end{equation}
where $\cL$ is the Lagrangian, $L_\xi$ denotes the Lie derivative in
the direction $\xi$ and contraction $I_\xi$ (acting on a $p$-form $\alpha$) is
defined to be
$$
I_\xi \alpha_p=\frac{1}{(p-1)!}\xi^\mu\,
\alpha_{\mu\nu^1...\nu^{p-1}}dx^{\nu^1}\wedge\cdots\wedge
dx^{\nu^{p-1}}\,.
$$
By direct calculation we find
$$  \begin{aligned}
 16\pi\,   J[\xi] &= \left(F_{IJ}-\beta B_{IJ}-\frac{\alpha}{2}B^{KL} \,\epsilon_{IJKL4}\right)\wedge I_\xi B_{IJ}\\
&+I_\xi A_{IJ}\wedge\left(\cD^A B^{IJ}\right)  +d\left(B^{IJ}\wedge
I_\xi A_{IJ}\right)\,.
 \end{aligned}$$
When field equations are satisfied this current is an exact
differential of a two form and thus we can write down the associated
charge to be
\begin{equation}\label{19}
    Q[\xi]=\frac1{16\pi}\int_{\partial \Sigma} B^{IJ}\, I_\xi A_{IJ}
\end{equation}
which, after substituting the solution of the $B$ field equations
takes the form
\begin{equation}\label{20}
     Q=\frac1{16\pi}\int_{\partial \Sigma} \left(\frac{1}{2} M^{ab}{}_{cd}\, F_{ab}\, I_\xi \omega^{cd}-\frac{2}{\beta\ell^2}\,
     T_a \, I_\xi e^a\right)\,,
\end{equation}
where $\partial\Sigma$ is a spatial section of the manifold.

One can  check that the expression for the Noether charge (\ref{20})
agrees with the one that can be obtained from the first order action
(\ref{our}), as it should. It is also worth noticing that the
Noether charge can be expressed compactly as
$$
Q=\frac{1}{16\pi}\int_{\partial \Sigma} \frac{\delta \cL}{\delta F^{IJ}}\;I_\xi A^{IJ}\,.
$$

Turning back to the formula (\ref{20}) and taking torsion \mbox{$T^a=0$} we
can express the charge in the final form
\begin{equation}\label{21}
   Q[\xi]=\frac{\ell^2}{32\pi G}\int_{\partial \Sigma} I_\xi \omega_{ab}
   \left(\epsilon^{ab}_{\;\;\;cd}F^{cd}_{jk}-2\gamma F^{ab}_{jk}\right)dx^j\wedge dx^k\,.
\end{equation}
This generalizes the result of \cite{Wald:1993nt},
\cite{Iyer:1995kg}, \cite{Aros:1999kt}, \cite{Aros:1999id}, and
\cite{Olea:2005gb} to the case of first order gravity with Immirzi
parameter.

Having the general expression for the charge, we can now turn to
finding the formula for the entropy. According to \cite{Wald:1993nt} and
\cite{Iyer:1995kg} the black hole entropy $S$ is proportional to the
value of the Noether charge (\ref{21}) calculated at the black hole
horizon and associated with a timelike Killing vector
$\partial/\partial t$, which vanishes at the horizon $\partial\Sigma_H$
\begin{equation}\label{22}
   \left. Q\left(\frac\partial{\partial t}\right)\right|_{\partial\Sigma_H} = \frac{\kappa}{2\pi}\,
   \mbox{Entropy}\, ,
\end{equation}
where $\kappa$ is the surface gravity. The question we would like to
address here is how the presence of the Immirzi parameter influence
the resulting expression for entropy. In this paper we will
investigate only the case of AdS--Schwarzschild black hole, leaving
another examples of the asymptotically anti-de Sitter black hole
spacetimes to the forthcoming publication.\newline

To calculate the value of the Noether charges (\ref{21}) for the
Schwarzschild--AdS spacetime let us first fix the metric to be
\begin{equation}\label{23}
    ds^2=-f(r)^2 dt^2 +f(r)^{-2} dr^2+r^2(d\theta^2+\sin^2 \theta d\varphi^2)
\end{equation}
with
\begin{equation}\label{24}
    f(r)^2=(1-\frac{2GM}{r}+\frac{r^2}{\ell^2})\,.
\end{equation}
It can be checked that for the case of the metric (\ref{23}) the
surface gravity $\kappa$ defined by the equation
\begin{equation}
    I_\xi \omega^{ab} \xi_b=\kappa\xi^a
\end{equation}
 is given by
 \begin{equation}\label{sg}
\kappa=\omega^{01}_t \Big|_{r_H}= \left(\frac{1}{2} \frac{\partial f(r)^2}{\partial r}
\right)\Big|_{r_H} \qquad\qquad T=\frac{\kappa}{2\pi}\,.
\end{equation}

The charge associated with the timelike Killing vector
$\xi \equiv\partial/\partial t$ equals
$$
   Q[\xi]=\frac{4\ell^2}{32\pi G}\int_{\partial \Sigma}  \omega^{01}_{t}\left(\epsilon_{0123}F^{23}_{jk}-\gamma F^{}_{jk\,01}\right)dx^j\wedge dx^k
=$$\begin{equation}\label{25} \frac{4\ell^2}{32\pi G}\int_{\partial \Sigma}
\left(\frac{1}{2} \frac{\partial f(r)^2}{\partial r}
\right)\left(1+\frac{r^2}{\ell^2}-f(r)^2\right)\sin \theta d\theta
\wedge d\varphi\,.
\end{equation}
Notice that this expression does not depend of the Immirzi
parameter; the $\gamma$-dependent terms in  (\ref{21}) have just
dropped out.

The value of this charge calculated at the boundary at infinity
gives
\begin{equation}\label{26}
    Q [\xi ]{}_\infty=\lim_{r\rightarrow\infty}\frac{1}{4\pi }
    \int_{\partial \Sigma}  \left( M+\frac{\ell^2 GM^2}{r^3}\right)
    \sin \theta d\theta \wedge d\varphi = M
\end{equation}
as it should be \cite{Aros:1999kt}, \cite{Aros:1999id}.\newline

The charge calculated at the Schwarzschild--AdS black hole horizon
equals
\begin{eqnarray}
    \label{27}
 Q[\xi]_H&=&\frac{\kappa \,\ell^2}{8\pi G}\left(1+\frac{r_H^2}{\ell^2}\right)\int_{\partial \Sigma}
\sin \theta d\theta \wedge d\varphi\nonumber\\
&=&\frac{\kappa}{2\pi}\frac{4\pi(r_H^2 +\ell^2)}{4G}\, ,
\end{eqnarray}
where $\kappa$ is the surface gravity defined by eq.\ (\ref{sg}).
The horizon radius $r_H$ is the largest real solution of the third
order equation
$$
  r^3/\ell^2+r-2GM=0\, ,
  $$
which allows us to rewrite the expression (\ref{27}) as
\begin{equation}\label{28}
   Q[\xi]_{H}= \kappa \;\frac{M\ell^2}{r_H}\,.
\end{equation}
From (\ref{27}) it's straightforward to see that the black  hole entropy yields the form
\begin{equation}\label{30}
    \textsc{S}=\frac{A}{4G}+\frac{4\pi\ell^2}{4G}\, .
\end{equation}
The first term is the standard Bekenstein--Hawking area law, while
the second is just a constant, which does not alter the
first law of thermodynamics. In the
first order formalism its presence can be regarded as price that has
to be paid for the regularization at infinity and the presence of
the Euler term in the action. The appearance of the additive
constant in the expression for  black hole has been discussed in the
context of Lovelock and Gauss-Bonet gravity theories e.g.\ in
\cite{Krishnan:2009tj}, \cite{Sarkar:2010xp}. It is worth mentioning that our model avoids the problem of the negative entropy \cite{Clunan:2004tb}.

In the context of the BF construction presented above this constant
could be understood as an indication that the vacuum of the
constrained BF theory (being the maximally symmetric spacetime with
$SO(2,3)$ symmetry) carries some entropy. A deeper origin of this
entropy remains still to be understood.

\section{Discussion: what about the Immirzi parameter?}

In the precedent section we calculated the entropy of the
Schwarzschild--AdS black hole by making use of the Wald-Iyer
prescription. Interestingly, the resulting expression (\ref{30})
does not contain any trace of the Immirzi parameter, in spite of the
fact this parameter was present in the Lagrangian of the dynamical
theory that we started with.

Before turning to the discussion of this intriguing result let us
try to trace the reason for the Immirzi parameter disappearance. Let
us consider, for simplicity, the action without the regularizing
Euler and Pontryagin terms, using just the Palatini and Holst
actions. In the case of axisymmetric stationary spacetime the charge
associated with the Killing vector $\partial_\chi$ being either
$\partial_t$ or $\partial_\varphi$ (related to the mass and angular
momentum at infinity) reads
\begin{eqnarray}
Q[\partial_\chi]&=&\frac{1}{32\pi G}\int_{\partial\Sigma}\omega^{ab}_\chi
\big(\epsilon_{abcd}(e^c_\theta e^d_\varphi-e^c_\varphi e^d_\theta)\nonumber\\
&-&\frac{2\gamma}{32\pi G}\int_{\partial\Sigma}\omega^{ab}_\chi
\left(e_{\theta\,a}e_{\varphi\,b}-e_{\varphi\,a}e_{\theta\,b})\right)\,.
 \end{eqnarray}
Using the definition of  the connection
$\omega^{ab}_\mu=e^{\nu\,a}\nabla_\mu
e^b_\nu=e^{\nu\,a}\left(\partial_\mu
e^b_\nu-\Gamma^\lambda_{\;\mu\nu}e^{b}_\lambda \right)$ we can
drastically simplify this formula  and for $\chi=t$ we have
\begin{equation}
   Q[\partial_t]=\frac{1}{16\pi G}\int_{\partial\Sigma} \left(\epsilon^{\mu\nu}_{\quad\theta\varphi}\Gamma_{\mu t\nu}-\gamma\,(\Gamma_{\theta t\varphi}-\Gamma_{\varphi t\theta})\right)\,.
\end{equation}
Therefore Immirzi parameter might be present in the expression for
black black hole thermodynamics if
\begin{equation}
  \gamma  \int_{\partial \Sigma}\partial_\theta g_{t\varphi}\neq 0\,.
\end{equation}
Thus we expect that such a contribution proportional to Immirzi
parameter can be present in Taub--NUT--AdS spacetime, and it is
going to be proportional to Taub--NUT 'mass', and not to the
Schwarzschild one. We will present the detailed discussion of
several black hole asymptotically AdS spacetimes in the forthcoming
paper.

Let us now return to the problem if and how our expression for the
entropy (\ref{30}) can be reconciled with the Loop Quantum Gravity
calculation \cite{Ashtekar:2000eq}, \cite{Domagala:2004jt},
\cite{Meissner:2004ju}, \cite{Agullo:2010zz} according to which the black hole entropy computed by counting the black hole horizon microstates equals
\begin{equation}\label{31}
\textsc{S}_{LQG}=\frac{\gamma_M}{\gamma}\frac{A}{4G}
\end{equation}
where $\gamma_M$ is a parameter, whose numerical value is between
$0.2$ and $0.3$, accompanied by higher order corrections; see
\cite{Agullo:2010zz} for detailed discussion. It is not hard to
understand why  $\gamma$ should be explicitly present in this
formula. Indeed the Immirzi parameter defines the size of the
quantum of the area, and therefore it must show up in the state
counting for black hole horizon. It would have been for some quite
unnatural cancelations to make it disappear from the entropy formula
in the semiclassical limit of Loop Quantum Gravity. Yet the
expression for entropy presented above (\ref{30}), which holds in
the semiclassical theory, whose quantum counterpart LQG is supposed
to be, shows no trace of $\gamma$.

A possible way to resolve this dilemma, as suggested in
\cite{Jacobson:2007uj}, is to notice that the entropy in (\ref{31})
was calculated using microscopic quantities, while in eq.\
(\ref{30}) with the help of those of effective low energy ones. It
follows that there might be highly nontrivial relations between the
area $A$ and and Newton's constant $G$ of (\ref{31}) and those of
(\ref{30}), so that, when the relations between them are properly
understood, and the renormalization effects are taken into account,
the two expression may turn out to be completely equivalent.

Another possible way out  was proposed recently in
\cite{Perez:2010pq}. In this paper it was observed that there exist
an additional ambiguity parameter associated with the construction
of the $SU(2)$ Chern--Simons theory that describes the microscopic
degrees of freedom of the isolated horizon. This parameter is of the
similar nature as the Immirzi one, and one can adjust the two in
such a way, so as to make the final expression for the black hole
entropy having the standard Bekenstein--Hawking form.

In both cases it remains to be understood in details how the
proposed mechanisms work. This question is related to the notorious
problem of the semiclassical limit of Loop Quantum Gravity, and it
seems that without controlling this limit one cannot make any
definite conclusions.

\section*{ACKNOWLEDGEMENTS}
We thank A.\ Perez for discussion and bringing the paper
\cite{Perez:2010pq} to our attention.   The work of J.\
Kowalski-Glikman was supported in part by grants 182/N-QGG/2008/0,
and the work of R.\ Durka was supported by the National Science
Centre grant  N202 112740.

\end{document}